\documentclass[%
 reprint,
superscriptaddress,
 amsmath,amssymb,
 aps,
 prf,
]{revtex4-2}
\newcommand{\um}{\text{\textmu{}m}}
\newcommand{\etal}{\textit{et~al.}}
\usepackage{textcomp}
\usepackage{units}
\usepackage{hyperref}
\usepackage{graphicx}
\usepackage{dcolumn}
\usepackage{bm}

\begin{document}

\preprint{APS/123-QED}

\title{Cell-free layer development and spatial organization of healthy and rigid red blood cells in a microfluidic bifurcation}

\author{Yazdan Rashidi}
 \affiliation{Dynamics of Fluids, Department of Experimental Physics, Saarland University, Saarbr{\"u}cken, Germany}%
\author{Othmane Aouane}
 \affiliation{Helmholtz Institute Erlangen-N{\"u}rnberg for Renewable Energy, Forschungszentrum J{\"u}lich, Erlangen, Germany}%
 \author{Alexis C. Darras}
 \affiliation{Dynamics of Fluids, Department of Experimental Physics, Saarland University, Saarbr{\"u}cken, Germany}%
 \author{Thomas John}
 \affiliation{Dynamics of Fluids, Department of Experimental Physics, Saarland University, Saarbr{\"u}cken, Germany}%
 \author{Jens Harting}
 \affiliation{Helmholtz Institute Erlangen-N{\"u}rnberg for Renewable Energy, Forschungszentrum J{\"u}lich, Erlangen, Germany}%
 \affiliation{Department of Chemical and Biological Engineering and Department of Physics, Friedrich-Alexander-Universit{\"a}t Erlangen-N{\"u}rnberg, Erlangen, Germany}%
 \author{Christian Wagner}
 \affiliation{Dynamics of Fluids, Department of Experimental Physics, Saarland University, Saarbr{\"u}cken, Germany}%
 \affiliation{Physics and Materials Science Research Unit, University of Luxembourg, Luxembourg, Luxembourg}
 \author{Steffen M. Recktenwald}
 \email{steffen.recktenwald@uni-saarland.de}
 \affiliation{Dynamics of Fluids, Department of Experimental Physics, Saarland University, Saarbr{\"u}cken, Germany}
\date{\today}

\begin{abstract}
Bifurcations and branches in the microcirculation dramatically affect blood flow as they determine the spatiotemporal organization of red blood cells (RBCs). Such changes in vessel geometries can further influence the formation of a cell-free layer (CFL) close to the vessel walls. Biophysical cell properties, such as their deformability, which is impaired in various diseases, are often thought to impact blood flow and affect the distribution of flowing RBCs. This study investigates the flow behavior of healthy and artificially hardened RBCs in a bifurcating microfluidic T-junction. We determine the RBC distribution across the channel width at multiple positions before and after the bifurcation. Thus, we reveal distinct focusing profiles in the feeding mother channel for rigid and healthy RBCs that dramatically impact the cell organization in the successive daughter channels. Moreover, we experimentally show how the characteristic asymmetric CFLs in the daughter vessels develop along their flow direction. Complimentary numerical simulations indicate that the buildup of the CFL is faster for healthy than for rigid RBCs. Our results provide fundamental knowledge to understand the partitioning of rigid RBC as a model of cells with pathologically impaired deformability in complex in vitro networks.
\end{abstract}

\maketitle

\section{Introduction}
Blood circulation in the human body is a crucial process ensuring the delivery of oxygen and various nutrients to our organs through the circulatory system. The circulatory system is a complex network of bifurcating and branching vessels. Such bifurcating vessels can strongly affect the distribution and hematocrit partition of passing red blood cells (RBCs).\cite{Pries2008, Secomb2017} The rigidity of RBCs is often thought to impair blood flow and affect the spatiotemporal organization of flowing RBCs. Rigid RBCs with impaired deformability are found in multiple diseases, such as malaria, diabetes, sickle cell disease, or acanthocytosis.\cite{Stuart1990, Dondorp2002, Symeonidis2001, Mannino2012, Reichel2022, Rabe2021} However, the fundamental mechanisms through which rigid RBCs modify blood flow have not been characterized extensively.

One of the non-trivial features of RBC flow in the circulation is their partitioning through vessel bifurcations.\cite{Fung1971} Recently, many studies have focused on parameters affecting the deviation from the well-established empirical model from Pries~\etal\cite{Pries1989, Shen2016a}. Among them, several studies have pointed out the role of cell focusing and the development of the cell-free layer (CFL) along the daughter vessels.\cite{Mantegazza2020} The formation of a CFL is crucial for blood flow in vivo, as it reduces its hydrodynamic resistance.\cite{Fahraeus1929a, Fahraeus1931} Furthermore, it is important for various biomedical applications such as plasma separation, especially in complex geometries,\cite{Faivre2006, Kim2009, Xiang2015, Tripathi2015} and highly relevant for disease diagnostics of cells with impaired deformability.\cite{Claveria2021, Fedosov2010, Makena2011} Such an impaired RBC deformability severely impacts the viscosity and the shear-thinning behavior of blood and RBC suspensions, thus affecting hemorheology, flow resistance, and microvascular perfusion.\cite{Chien1967, Chien1987, Lanotte2016, Passos2019} Moreover, RBC deformability can influence the emergence of a CFL or cell-depleted zones in complex microfluidic geometries that are often used in lab-on-a-chip devices.\cite{Fujiwara2009, Shen2016a, Abay2020}

In general, the CFL in microfluidic devices depends on various factors, including hematocrit, channel dimensions, and flow rate.\cite{Kim2009, Tripathi2015} Moreover, changes in the CFL in dilute suspensions can also arise from geometric features of the channel, such as confinement,\cite{Kaoui2012, Tomaiuolo2012, Iss2019} constrictions,\cite{Faivre2006, Abay2020} and bifurcations.\cite{Sherwood2014, Shen2016a} Therefore, the formation of a CFL of healthy RBC suspensions in complex geometries, such as vessel networks has received increasing attention in recent experimental and numerical investigations.\cite{Shen2016a, Kaliviotis2017, Bento2018, Bento2019, Yamamoto2020, Zhou2021} Although the biophysical RBC properties, such as their deformability, were found to influence the CFL as well,\cite{Amini2014a, Abay2020} detailed knowledge about the effects of RBCs with impaired deformability on the spatiotemporal RBC organization and the partitioning in complex geometries remains scarce.

In this study, we explore how the rigidity of RBCs modifies their focusing and subsequent formation of a CFL in a bifurcating microfluidic channel. Therefore,  we artificially rigidify RBCs using glutaraldehyde and examine their spatial distribution at various positions in the microfluidic T-junction. We investigate  the evolution of the RBC distribution across the channel width, as well as the formation of a CFL along the mother and the subsequent daughter channels. Our results on rigid RBC are compared with investigations of healthy cells at the same RBC concentrations (\mbox{$\unit [0.1-5]{\%Ht}$}). We observe that the RBC rigidity drastically alters the RBC organization at the bifurcation after the mother channel, resulting in distinct lateral distributions of RBCs in the daughter channels. The magnitude of this effect increases with increasing inertia. Furthermore, we show how the CFL develops in the daughter vessels of the bifurcation. Our numerical simulations demonstrate that the distance that is required to reach a steady CFL in the straight daughter vessel is shorter for healthy than for rigid RBCs. Our work offers further insights into the flow behavior of RBCs with impaired deformability and how rigid cells can impair blood circulation.

\section{Materials and Methods}
\subsection{Experimental}
\subsubsection{Sample preparation}
Blood is taken with informed consent from healthy voluntary donors. It is suspended in phosphate-buffered saline (PBS) solution (Gibco PBS, Fisher Scientific, Schwerte, Germany) and centrifuged at \mbox{$\unit[1500]{g}$} for five minutes to separate RBCs and plasma. Sedimented RBCs are subsequently resuspended in PBS and the centrifugation and washing steps are repeated three times. For the final suspensions, hematocrit concentrations of \mbox{$\unit [0.1]{\%Ht}$}, \mbox{$\unit [1]{\%Ht}$}, and \mbox{$\unit [5]{\%Ht}$} are adjusted in a PBS solution that contains \mbox{$\unit [1]{g\,L^{-1}}$} bovine serum albumin (BSA, Sigma-Aldrich, Taufkirchen, Germany). Furthermore, samples with artificially rigidified RBCs are prepared. For this, washed RBCs are incubated in a \mbox{$\unit[0.1]{\%}$} glutaraldehyde (GA, grade I solution, Merck KGaA, Darmstadt, Germany) solution for one hour according to Abay~\etal\cite{Abay2019}. Subsequently, RBCs are washed with PBS to remove excess GA and are suspended in a PBS/BSA solution at the same concentrations as for healthy RBCs. 

Blood withdrawal, sample preparation, and experiments were performed according to the guidelines of the Declaration of Helsinki and approved by the ethics committee of the `{\"A}rztekammer des Saarlandes' (approval number 51/18).

\subsubsection{Microfluidic setup}
\begin{figure*}
\includegraphics[width=17.1cm]{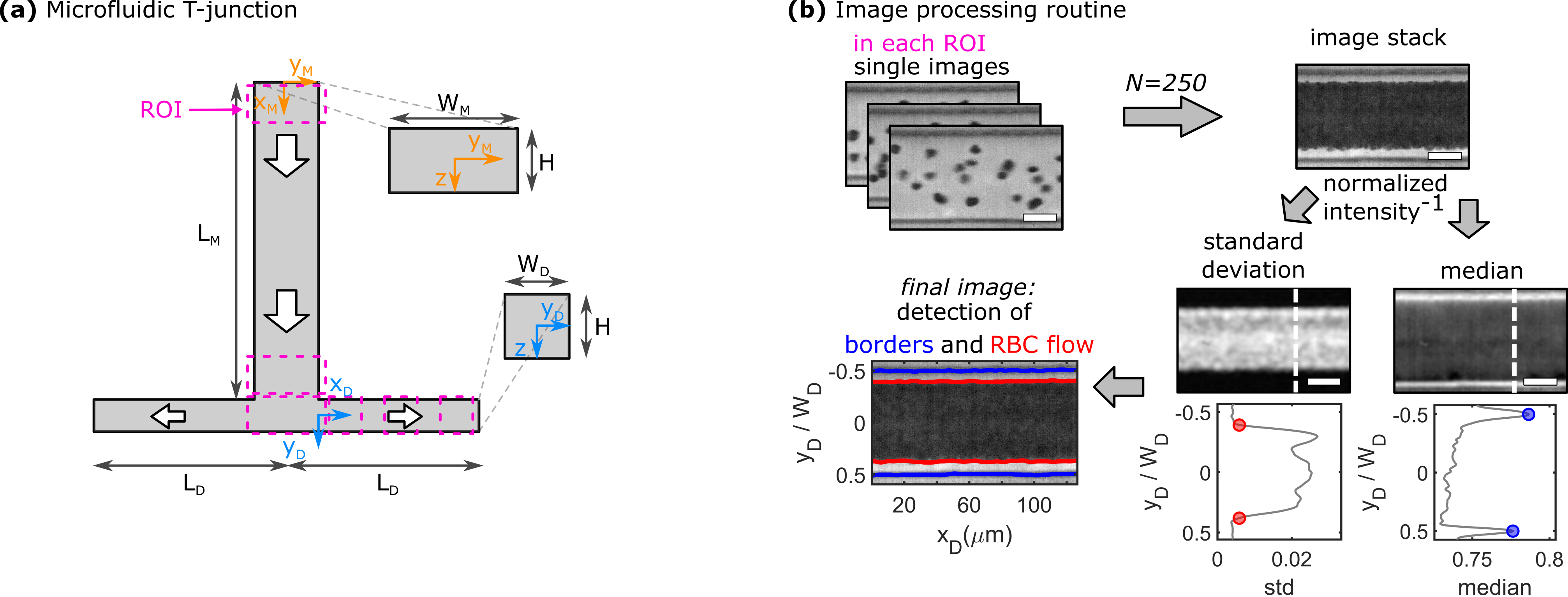}
\caption{Schematic representation of the microfluidic setup and the experimental data analysis routine. (\textbf{a}) T-junction geometry, consisting of an inlet mother channel with width \mbox{$W_{\text{M}}$}, height \mbox{$H$}, and length \mbox{$L_{\text{M}}$} and two outlet daughter channels each with width \mbox{$W_{\text{D}}$}, height \mbox{$H$}, and length \mbox{$L_{\text{D}}$}. Dashed magenta boxes correspond to different regions of interest (ROI) for data acquisition in the mother and daughter vessels. (\textbf{b}) Image processing routine to determine the CFL thickness. Scale bars represent \mbox{$\unit[20]{\um}$}. 250 single images are stacked and the standard deviation and median of the intensity are derived from the inverted image. Along each vertical pixel line, the intensity's standard deviation and the intensity's median are calculated. Two representative plots of the two parameters are shown below the images, corresponding to a horizontal position marked by the dashed white lines. Red dots correspond to a threshold value of \mbox{$0.5{\%}$} of the measured maximum inverted intensity to detect the RBC flow and blue dots highlight the peak position of the median intensity, which corresponds to the position of the walls. The red and blue lines in the final image correspond to the determined channel borders and borders of the RBC flow, respectively.}
\label{FIG_Setup}
\end{figure*}

The RBC suspensions are pumped through a microfluidic chip that contains a T-junction geometry. The microfluidic device is fabricated using polydimethylsiloxane (PDMS, RTV 615A/B, Momentive Performance Materials, Waterford, NY) through standard soft lithography.\cite{Friend2010} The T-channel has a height of \mbox{$H=\unit[54\pm 1]{\um}$} in $z$-direction and consists of an inlet channel, referred to as mother channel, with a width of \mbox{$W_{\text{M}}=\unit[104\pm 2]{\um}$}, and a length  \mbox{$L_{\text{M}}=\unit[3]{cm}$} in the flow direction. At the T-bifurcation, the mother channel splits into two daughter outlet channels, each with a width of \mbox{$W_{\text{D}}=\unit[55\pm 1]{\um}$} and a length of \mbox{$L_{\text{D}}=\unit[1.75]{cm}$}. Figure~\ref{FIG_Setup}a shows a schematic representation of the microfluidic T-junction and the used coordinate systems. The mother and daughter channels are connected with rigid medical-grade polyethylene tubing (\mbox{$\unit[0.86]{mm}$} inner diameter, Scientific Commodities, Lake Havasu City, AZ) to the sample and waste containers, respectively. The microfluidic device is mounted on an inverted microscope (Eclipse TE2000-S, Nikon, Melville, New York), equipped with a LED illumination and a \mbox{$40\times$} air objective (Plan Fluor, Nikon, Melville, NY) with a numerical aperture \mbox{$NA=0.6$}. A high-precision pressure device (OB1-MK3, Elveflow, Paris, France) is used to pump the suspensions through the channel at constant pressure drops of \mbox{$p = \unit[200]{mbar}$}, \mbox{$\unit[400]{mbar}$}, \mbox{$\unit[600]{mbar}$}, and \mbox{$\unit[800]{mbar}$}. 

Microfluidic experiments are performed both with symmetric as well as asymmetric flow rates in both daughter vessels. For the symmetric case, both daughters have the same length \mbox{$L_{\text{D}}$}. To introduce an asymmetric flow rate, the length of one daughter vessel is shortened to \mbox{$L_{\text{D}}^*=3L_{\text{D}}/4$}, \mbox{$L_{\text{D}}/2$}, or \mbox{$L_{\text{D}}/4$}, while the length of the other daughter channels is kept constant at \mbox{$L_{\text{D}}$}. Shortening the length of one daughter channel decreases the hydraulic resistance in the channel, hence, leading to an increase in the flow rate in the short daughter vessel. We define the asymmetry ratio as \mbox{$\text{AR}=L_{\text{D}}/L_{\text{D}}^*$}, which is one in case of a symmetric bifurcation. 

\subsubsection{Data acquisition and analysis}
To determine the evolution of the RBC distribution across the channel width and to detect the CFL, the RBC flow is recorded at different regions of interest (ROIs) along the flow direction. Magenta boxes in Fig.~\ref{FIG_Setup}a indicate ROIs corresponding to two positions at the beginning and end of the mother channel before the bifurcation, respectively, one at the T-bifurcation, as well as three positions in the daughter channels after the bifurcation. 

At low hematocrit (\mbox{$\leq \unit [1]{\%Ht}$}), a high-speed camera (MEMRECAM GX1, NAC Image Technology, Salem, MA) with a frame rate of \mbox{$\unit[20,000]{fps}$} (frames per seconds) is used, which allows us to resolve the trajectories and determine the velocities of individual RBCs by means of particle-tracking velocimetry (PTV). Therefore, the positions of individual RBCs are determined in each frame and linked in consecutive frames to calculate the RBC trajectories using a self-written \textsc{Matlab} (The MathWorks, Natick, MA) PTV algorithm. Based on those trajectories, the velocity profile and the RBC distribution across the channel width in $y$-direction are calculated.

To assess the relative influence of inertial forces, we calculate the Reynolds number \mbox{$\text{Re}$}, which relates the inertial to viscous forces in the system. Here, we define it as \mbox{$\text{Re} = {u_{\text{max}} D_{\mathrm{h}}\rho}/{\eta}$}, with the fluid density \mbox{$\rho=\unit[1]{g\,cm^{-3}}$}, the maximum cell velocity \mbox{$u_{\text{max}}$} in the channel center, the fluid's dynamic viscosity \mbox{$\eta=\unit[1.2]{mPa\,s}$}, and the hydraulic diameter of the rectangular microfluidic channel \mbox{$D_{\mathrm{h}}=2W_{\mathrm{M}} H/(W_{\mathrm{M}}+H)$}. Representative plots of the velocity profiles in the mother and daughter channels are shown in Fig.~S1 in the Supplementary Material. Based on the applied pressure drop and the resulting maximum cell velocity \mbox{$u_{\text{max}}$} (see Fig.~S2 in the Supplementary Material), the Reynolds number is always greater than one (\mbox{$\text{Re}=6-24$}), hence inertial forces cannot be neglected.

At higher RBC concentrations, individual cells cannot be discriminated and the cell distribution across the channel width and the CFL cannot be determined based on the RBC trajectories. Therefore, RBC flow is recorded with a high-speed camera (Fastec HiSpec 2G, FASTEC Imaging, San Diego, CA) at a lower frame rate of \mbox{$\unit[1,500]{fps}$}. A self-written \textsc{Matlab} program is used to determine the channel borders, the core RBC flow, and thus, the thickness of the CFL, as exemplified in Fig.~\ref{FIG_Setup}b. First, 250 images of the recorded sequence are  inverted and then stacked. Second, the pixel-wise standard deviation and median of the intensity in the stack are calculated. Regions with a high standard deviation correspond to locations in the channel where the image intensity fluctuates the most, hence the core RBC flow. The median of the intensity is most prominent at the channel borders, which do not change their position over the image sequence. Based on the peak positions of the median inverted intensity and a threshold of \mbox{$0.5{\%}$} of the measured maximum inverted intensity, the channel borders and the region of the RBC flow are determined, respectively. Subsequently, the thickness of the CFL is calculated for each vertical pixel line. A representative depiction of the resulting final image of the border and core RBC flow detection is shown in the bottom left of Fig.~\ref{FIG_Setup}b. Experimental CFL results are shown as mean values with corresponding standard deviations between different measurements as error bars.

\subsection{Simulations}
We simulate the fluid motion in three dimensions using the standard D3Q19 lattice Boltzmann method (LBM) \cite{benzi1992lattice, succi2001} with the Bhatnagar-Gross-Krook (BGK) collision operator.\cite{bhathnagor1954model} The RBCs are modeled as hyperelastic biconcave capsules endowed with shear and bending resistance. We use the Skalak strain model to account for the RBC membrane's shear elasticity and local area dilatation.\cite{skalak1973strain,skalak1973modelling} The resulting elastic force is computed using a linear finite element method (FEM).\cite{kruger2011efficient} The Skalak model has been initially tailored for RBCs and used successfully to recover their dynamics and morphologies under different flow conditions.\cite{omori2014numerical, gross2014rheology, sinha2015dynamics, Aouane2021a} The bending is accounted for using the Helfrich free energy developed for bilayer lipid membranes.\cite{helfrich1973elastic} The bending force is obtained through the functional derivative of the Helfrich free energy, and its terms are numerically evaluated using discrete differential geometry operators for triangulated meshes.\cite{meyer2003discrete, sinha2015dynamics} The impermeability and inextensibility of the RBC membrane are fulfilled by imposing penalty functions on both the global volume and area.\cite{gross2014rheology} The coupling between the fluid and the particles is based on the immersed boundary method (IBM).\cite{peskin2002immersed} Additional details on the numerical method are provided in appendix A at the end of the manuscript. 

Simulations are performed in a straight channel with the same dimensions as the daughter vessels in the microfluidic experiments to determine the steady-state behavior of the CFL. Since the magnitude of the CFL decreases with increasing RBC concentration, we limit our numerical simulations to the lowest hematocrit of \mbox{$\unit[0.1]{\%Ht}$} that was tested experimentally.  In the numerical simulations, the CFL is calculated based on the trajectories of the RBCs' center of mass. Since the RBCs have a spatial extension between \mbox{$\unit[2]{\um}$} and \mbox{$\unit[4]{\um}$} around their center of mass, the CFL obtained from the numerical simulation should be reduced by a value between \mbox{$\unit[2]{\um}$} and \mbox{$\unit[4]{\um}$} to be compared with experimental measurements. Hence, numerical data of the CFL is shown as the cell's center of mass \mbox{$+\unit[3]{\um}$} and with an error of \mbox{$\pm\unit[1]{\um}$}.

\section{Results}
\subsection{RBC focusing and distribution at low concentrations in the microfluidic T-junction}

\begin{figure*}
\centering
\includegraphics[width=17.1cm]{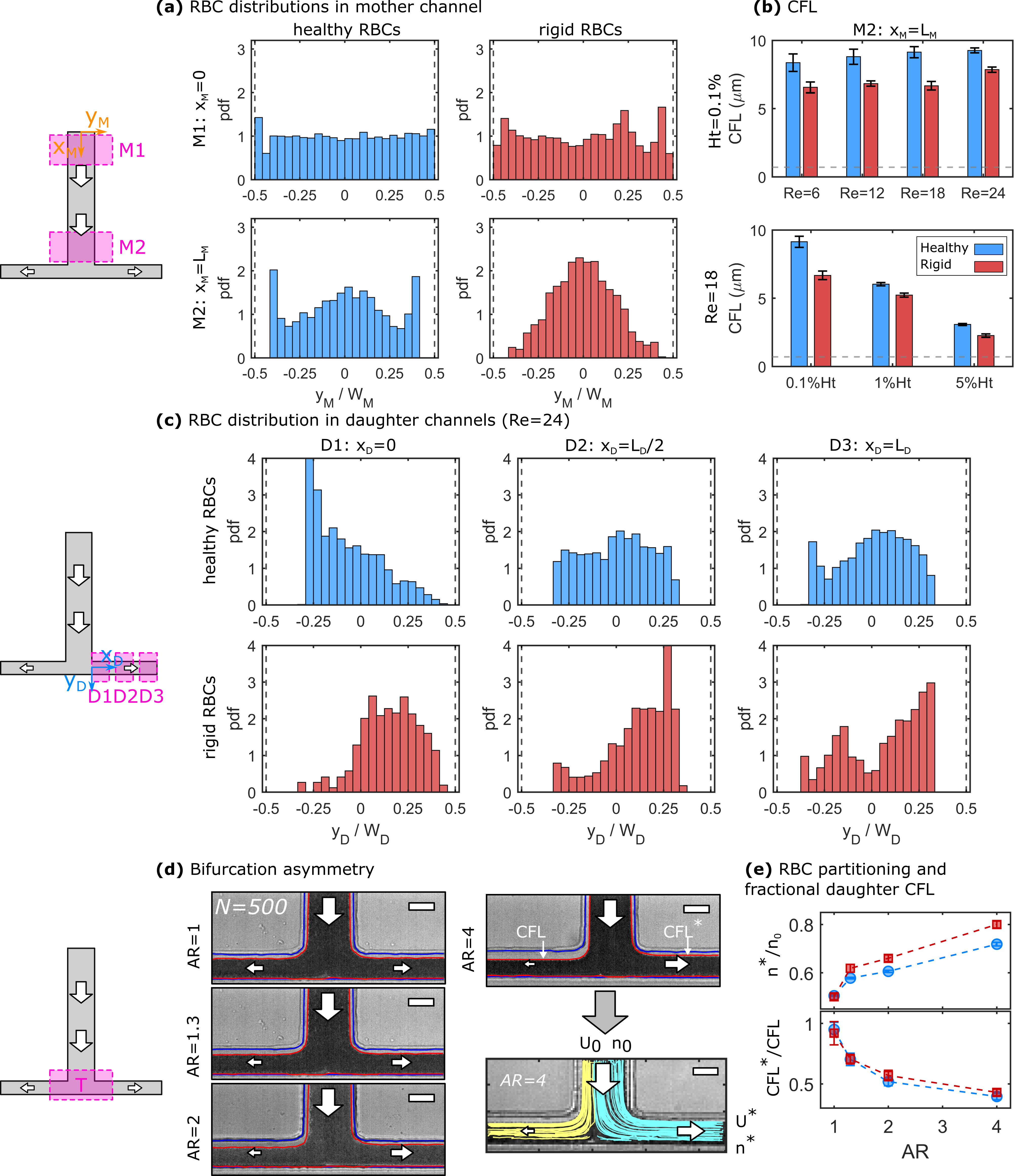}
\caption{Flow of a \mbox{$\unit [0.1]{\%Ht}$} RBC suspensions through the microfluidic T-junction. (\textbf{a}) Probability density distributions (pdf) for healthy (left) and rigid (right) RBCs at the beginning (top) and end (bottom) of the mother channel at \mbox{$\text{Re}=24$}. Dashed vertical lines indicate the position of the channel borders. (\textbf{b}) CFL at the end of the mother channel as a function of \mbox{$\text{Re}$} at \mbox{$\unit [0.1]{\%Ht}$} (top) and the RBC concentration at \mbox{$\text{Re}=18$} (bottom). Horizontal lines correspond to the optical resolution limit of two pixels.  (\textbf{c}) Distribution of healthy (top) and rigid (bottom) RBCs at three positions along the flow direction in the daughter vessels for the symmetric partitioning at \mbox{$\text{Re}=24$}. (\textbf{d}) Stacks of 500 images showing RBC partitioning at the bifurcation for different asymmetry ratios \mbox{$\text{AR}$}. The red and blue lines in the image stack correspond to the channel borders and the RBC flow, respectively. For \mbox{$\text{AR}=4$}, representative trajectories are shown. Scale bars represent \mbox{$\unit[50]{\um}$}. (\textbf{e}) Relative RBC flux in the shorter daughter vessel \mbox{$n^*/n_0$} (top) and the ratio of the upper CFL between the short and the long daughter \mbox{$\text{CFL}^*/\text{CFL}$} as a function of \mbox{$\text{AR}$}.}
\label{FIG_lowHt}
\end{figure*}

At finite inertia (\mbox{$\text{Re}>1$}), particles and cells in dilute suspensions can show ordering and flow focusing in straight vessels.\cite{DiCarlo2009} In the mother channel, we observe a focusing of RBCs along the flow direction, as shown in Fig.~\ref{FIG_lowHt}a for healthy and rigid RBC at \mbox{$\unit [0.1]{\%Ht}$} and \mbox{$\text{Re}=24$}. The top and bottom panels in Fig.~\ref{FIG_lowHt}a correspond to a $x$-position close to the beginning (\mbox{$x_{\text{M}}\approx 0$}) and end (\mbox{$x_{\text{M}}= L_{\text{M}}$}) of the mother channel, respectively. At the beginning of the mother vessel (\mbox{$x_{\text{M}}\approx 0$}, Fig.~\ref{FIG_lowHt}a top), cells are homogeneously distributed across the channel width for both healthy and rigid RBCs. Furthermore, RBCs flow in the close vicinity of the channel walls. Hence, we do not experimentally observe a CFL at \mbox{$x_{\text{M}}\approx 0$}, indicated by the histogram bars at  \mbox{$y_{\text{M}}/W_{\text{M}}=\pm 0.5$}. However, at the end of the mother vessel (\mbox{$x_{\text{M}}=L_{\text{M}}$}), we find pronounced differences in the distributions for healthy and rigid RBCs. Healthy cells are focused on three equilibrium positions, namely one in the channel center and two close to the borders. Rigid RBCs preferentially flow in the center of the channel. Furthermore, both healthy and rigid RBCs exhibit a pronounced CFL close to the channel walls at \mbox{$x_{\text{M}}=L_{\text{M}}$}. The observed focusing effect towards distinct positions for healthy and rigid cells emerges for \mbox{$\text{Re}\geq12$} and its magnitude increases with increasing \mbox{$\text{Re}$}. Complementary results for other investigated \mbox{$\text{Re}$} are shown in Fig.~S3 in the Supplementary Material. 

Based on the RBC distributions across the channel width and the positions of the vessel borders, the CFL thickness is determined at the end of the mother vessel \mbox{$x_{\text{M}}=L_{\text{M}}$}. Figure~\ref{FIG_lowHt}b shows the dependency of the CFL for both rigid and healthy cells as a function of \mbox{$\text{Re}$} at \mbox{$\unit [0.1]{\%Ht}$} (top) and as a function of the RBC concentration at a constant \mbox{$\text{Re=18}$} (bottom). Within the investigated pressure drop range, the CFL only increases slightly for both rigid and healthy cells with increasing \mbox{$\text{Re}$}. However, we observe a strong decrease in the CFL thickness with increasing RBC concentration, as shown in the bottom panel of Fig.~\ref{FIG_lowHt}b. In general, rigid RBCs form a smaller CFL compared to healthy RBCs, in agreement with previous studies.\cite{Kim2009} This difference is most pronounced at low RBC concentrations.

At the end of the mother channel, the RBCs reach the T-bifurcation, where the flow splits into the two daughter branches. Similar to the mother channel, we perform RBC tracking at the beginning (D1: \mbox{$x_{\text{D}}\approx0$}), the middle (D2: \mbox{$x_{\text{D}}\approx L_{\text{D}}/2$}), and the end (D3: \mbox{$x_{\text{D}}\approx L_{\text{D}}$}) of the daughter branches. For the sake of simplicity and based on the top view projection of the T-junction, we refer to positions at \mbox{$-0.5 \leq y_{\text{D}}/W_{\text{D}} < 0$} as up and to positions at \mbox{$0 \leq y_{\text{D}}/W_{\text{D}} \leq 0.5$} as down. Hence, the upper CFL in the daughter branches corresponds to the continuation of the CFL in the mother channel. Figure~\ref{FIG_lowHt}c shows the RBC distribution across the daughter channel width for healthy (top) and rigid (bottom) RBCs after the symmetric bifurcation. Right after the T-bifurcation in D1, we observe a strong accumulation of healthy RBCs at an off-centered position at \mbox{$y_{\text{D}}/W_{\text{D}}\approx-0.28$} close to the upper channel wall of the daughter vessel. With increasing \mbox{$y_{\text{D}}/W_{\text{D}}$}, the local cell concentration at this position continuously decreases. Note that in D1, RBC flow in close proximity to the down wall at \mbox{$y_{\text{D}}/W_{\text{D}}=0.5$}, while a large upper CFL (\mbox{$-0.5\leq y_{\text{D}}/W_{\text{D}}\leq-0.3$})  is formed. As the RBCs flow through the daughter vessels, the peak in the distribution diminished and a more uniform distribution is found in D2 and D3 for the healthy RBCs. Furthermore, a CFL appears at \mbox{$0.3\leq y_{\text{D}}/W_{\text{D}}\leq0.5$} as the RBCs flow in the daughter branches. 

In contrast to the healthy cells, the distribution for rigid RBCs exhibits a broad off-centered peak around \mbox{$y_{\text{D}}/W_{\text{D}}\approx0.2$} right after the bifurcation in D1, as shown in the bottom panel of Fig.~\ref{FIG_lowHt}c. While flowing along the daughter channel, more rigid RBCs migrate towards the upper half of the channel (\mbox{$-0.5\leq y_{\text{D}}/W_{\text{D}}\leq 0$}). However, the peak in the distribution at \mbox{$y_{\text{D}}/W_{\text{D}}\approx0.2$} persists for rigid RBCs until the exit of the daughter vessel at D3. With increasing \mbox{$\text{Re}$}, it seems that these distinct focusing effects for healthy and rigid RBCs increase, as shown for other investigated \mbox{$\text{Re}$} in Fig.~S4 in the Supplementary Material. 

We further examine how an asymmetric bifurcating flow affects the RBC partitioning and the CFLs at the bifurcation. Figure~\ref{FIG_lowHt}d shows representative image stacks for the bifurcating flow of a \mbox{$\unit [0.1]{\%Ht}$} healthy RBC suspension at different asymmetry ratios \mbox{$\text{AR}$}. This asymmetry leads to different flow rates as well as to different CFLs in both daughter vessels. To assess the relative RBC partitioning and flux, individual RBCs are tracked at the T-junction, representatively shown for 90 trajectories at \mbox{$\text{AR}=4$} in Fig.~\ref{FIG_lowHt}d. The two colors represent trajectories that end in either the left or the right daughter channel. The fraction of RBCs that end in the shorter daughter vessel \mbox{$n^*$} with respect to the total number of RBCs that enter the bifurcation \mbox{$n_0$} is plotted as a function of the asymmetry ratio for healthy and rigid cells in the top panel of Fig.~\ref{FIG_lowHt}e. In the case of a symmetric bifurcation (\mbox{$\text{AR}=1$}), we find \mbox{$n^*/n_0=0.5$}. However, with increasing \mbox{$\text{AR}>1$}, the number of RBCs that flow into the short daughter channel increases. This increase seems to be more pronounced for rigid  than for healthy RBCs. Moreover, as visible from the snapshots in Fig.~\ref{FIG_lowHt}d, two distinct CFLs emerge close to the two channel borders in each daughter vessel. In Fig.~\ref{FIG_lowHt}c, we observe that the down CFL at the impacting wall is essentially zero, while the upper CFL closer to the mother channel is more pronounced. The ratio between the two upper CFLs of the short and long daughter \mbox{$\text{CFL}^*/\text{CFL}$} is shown in the bottom panel of Fig.~\ref{FIG_lowHt}d. Here, we find that with increasing \mbox{$\text{AR}$} the \mbox{$\text{CFL}^*/\text{CFL}$} ratio decreases.

\subsection{CFL development in symmetric daughter vessels}

\begin{figure*}
\centering
\includegraphics[width=\textwidth]{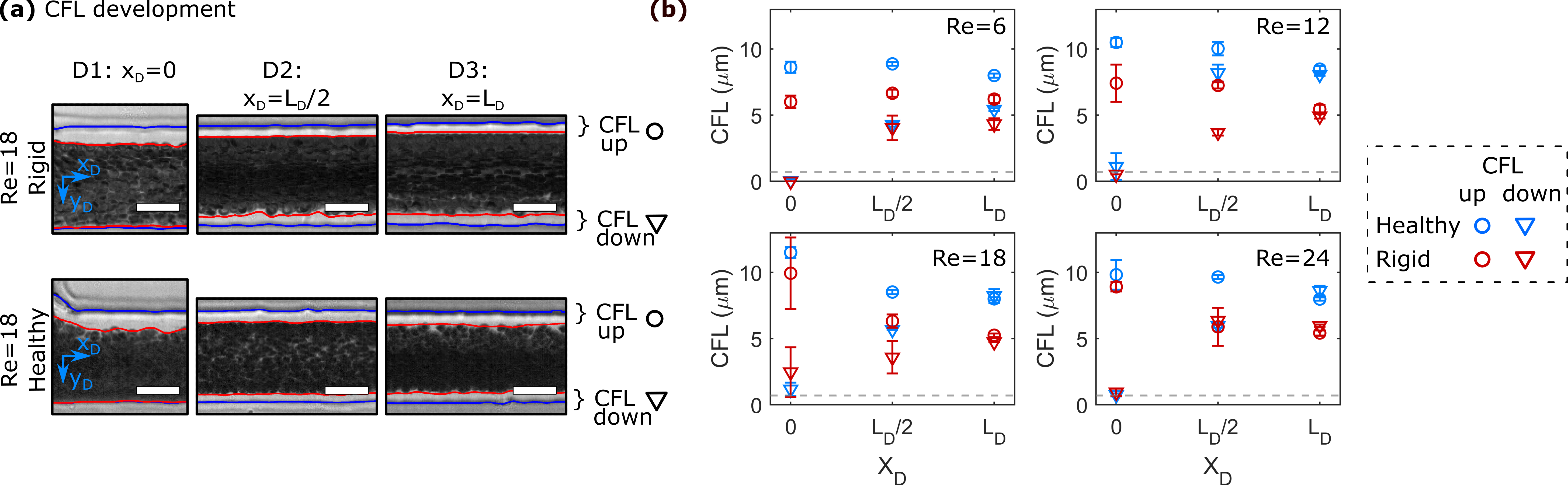}
\caption{CFL development along the flow direction in the daughter vessels. (\textbf{a}) Representative superimposed images for a \mbox{$\unit [0.1]{\%Ht}$} RBC suspension at three positions in the daughter channel at \mbox{$\text{Re}=18$} for (top) rigid and (bottom) healthy RBCs. The red and blue lines in the images correspond to the channel borders and the RBC flow, respectively. Scale bars represent \mbox{$\unit[20]{\um}$}. Up and down CFL correspond to the CFL in negative and positive $y_{\text{D}}$-direction, respectively. (\textbf{b}) Up and down CFL for a \mbox{$\unit [0.1]{\%Ht}$} RBC suspension as a function of $x_{\text{D}}$ for different \mbox{$\text{Re}$}.}
\label{FIG_CFLD}
\end{figure*}

To quantify the development of the CFL in the daughter vessels after the bifurcation in-depth, we employ the image processing routine shown in Fig.~\ref{FIG_Setup}b. Therefore, 50 images are stacked at each ROI and the position of the RBC core flow and the channel borders are determined. Representative images for the three positions along the daughter vessels are shown in Fig.~\ref{FIG_CFLD}a. Similar to the histogram-based analysis in Fig.~\ref{FIG_lowHt}, the notations up and down correspond to the CFL in negative and positive $y_{\text{D}}$-direction, respectively. 

Figure~\ref{FIG_CFLD}b shows the development of the asymmetric up and down CFLs along the flow direction in the daughter for different \mbox{$\text{Re}$}. At \mbox{$\text{Re}=6$}, the upper CFL is constant in the first two regions and slightly decreases at the end of the daughter vessel. Rigid RBCs show similar behavior as healthy RBCs at \mbox{$\text{Re}=6$}. However, they generate a smaller CFL compared to healthy cells. At $x_{\text{D}}=0$, the down CFL is essentially zero, since cells that flow in the center of the mother channel are pushed against the wall in the stagnation area of the bifurcation. As the cells travel along the daughter, the down CFL consistently builds up for both rigid and healthy RBCs. Similar to the upper CFL, the down CFL for healthy cells is larger than for rigid cells at the end of the daughter channel. However, for \mbox{$\text{Re=6}$}, the thickness of the down CFL does not seem to have reached the thickness of its upper counterpart, thus forming an asymmetric CFL at the end of the daughter branch at \mbox{$x_{\text{D}}=L_{\text{D}}$}.

Increasing \mbox{$\text{Re}$} affects the CFL development in the daughter vessel as follows:

(i) At the beginning of the daughter vessel at \mbox{$x_{\text{D}}=0$}, increasing \mbox{$\text{Re}$} leads to an increase in the upper CFL for both healthy and rigid RBCs. Due to the increased inertia, RBCs that initially flow in a central position of the mother channel are pushed into the T-junction wall opposite the mother channel. Therefore, the down CFL is still zero.  Moreover, RBCs that flow at streamlines close to the channel walls in the mother channel are pushed away from the wall to positions closer to the center of the daughter channel, thus forming a larger CFL directly after entering the daughter branch at \mbox{$x_{\text{D}}=0$}, reminiscent of cell-depleted zones generated by lip vortices in other complex flow fields of RBC suspensions.\cite{Brust2013}

(ii) In the second region of interest at \mbox{$x_{\text{D}}=L_{\text{D}}/2$}, the thickness of the upper CFL decreases for \mbox{$\text{Re} > 12$} compared to the upper CFL at the entrance of the daughter vessel. Opposite to this trend, the down CFL increases.

(iii) At the end of the daughter vessel at \mbox{$x_{\text{D}}=L_{\text{D}}$}, experiments at \mbox{$\text{Re}\geq 12$} result in the formation of an equal thickness in the up and down CFL. This symmetric CFL at \mbox{$x_{\text{D}}=L_{\text{D}}$} is independent of \mbox{$\text{Re}$} within the investigated regime.

Similar behavior regarding the CFL development in the daughter branches is also observed for higher RBC concentration, as shown in Fig.~S5 in the Supplementary Material. However, increasing the RBC concentration overall reduces the thickness of the CFL and results in a symmetric CFL already at \mbox{$\text{Re}=6$} for RBC suspension with \mbox{$\geq \unit [1]{\%Ht}$}.

\subsection{Assessment of the steady-state CFL through numerical simulations}

\begin{figure}
\centering
\includegraphics[width=8.3cm]{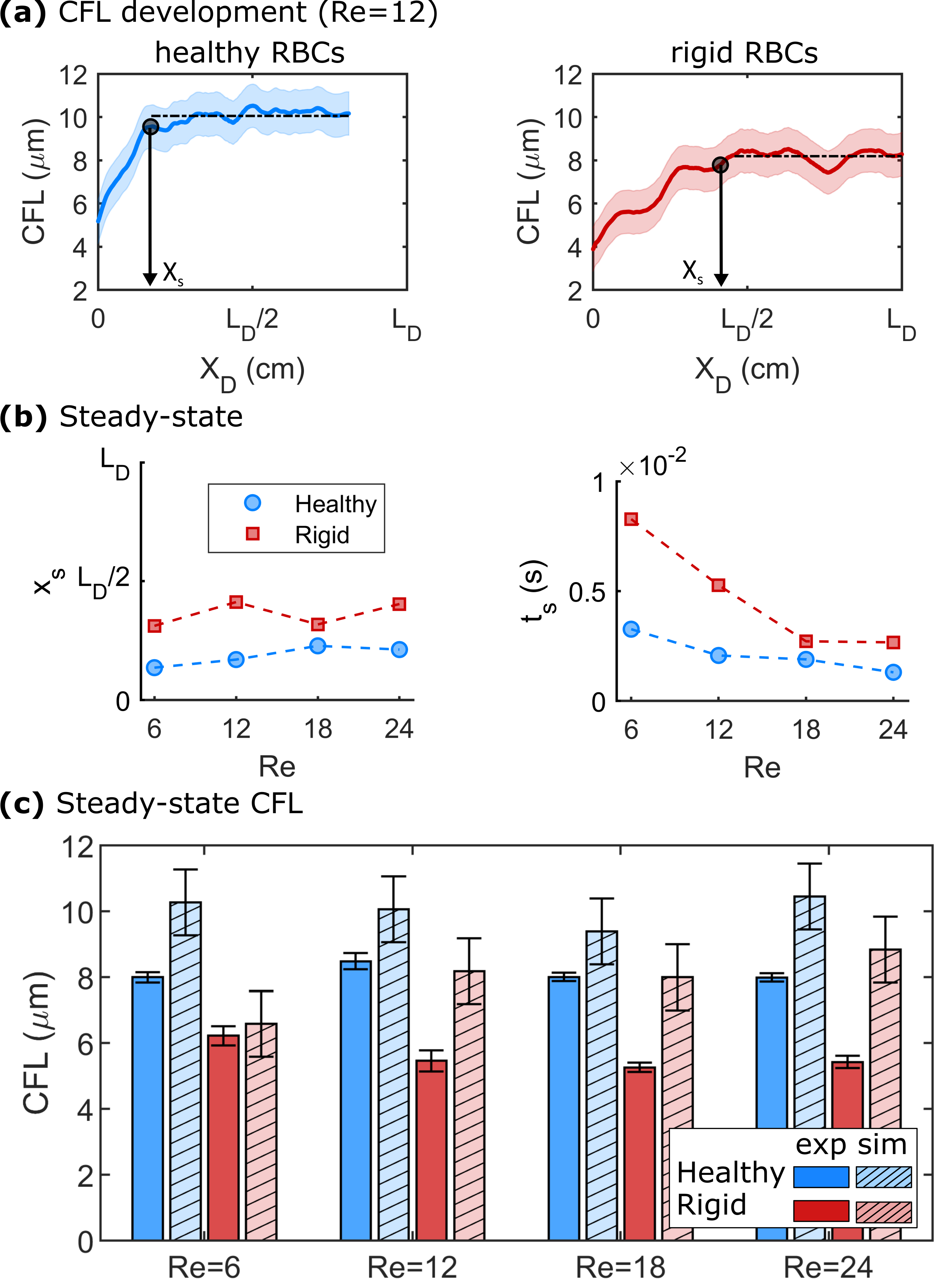}
\caption{Numerical simulations of healthy and rigid RBC suspension at \mbox{$\unit[0.1]{\%Ht}$} in a straight channel. (\textbf{a}) CFL development along the flow direction at \mbox{$\text{Re}=12$}. The $x$-axes in (\textbf{a}) are normalized by the length of the daughter branches \mbox{$L_{\text{D}}$} in the microfluidic experiments. The enveloping shaded areas correspond to \mbox{$\pm\unit[1]{\um}$}, as discussed in the method section. Black, dashed horizontal lines indicate the mean steady-state CFL and the black circles represent the distance \mbox{$x_s$} that the RBCs travel until the steady-state CFL is reached. (\textbf{b}) Length \mbox{$x_s$} (top) and corresponding time \mbox{$t_s$} (bottom) that required to achieve a steady-state CFL in the simulations as a function of \mbox{$\text{Re}$}. (\textbf{c}) Comparison between the experimental and numerical (hatched bars) results for different \mbox{$\text{Re}$}.}
\label{FIG_SIM}
\end{figure}

We further employ numerical simulations to determine the steady-state CFL of dilute healthy and rigid RBC suspensions at \mbox{$\text{Re}>1$}. Figure~\ref{FIG_SIM} highlights the results of the numerical simulations regarding the development of the CFL at \mbox{$\text{Re}>1$}. Representative developments of the CFL along the flow direction for healthy and rigid RBCs at \mbox{$\text{Re}=12$} are shown in Fig.~\ref{FIG_SIM}a. Here, the $x$-axes are normalized by the length of the daughter branches \mbox{$L_{\text{D}}$} in the microfluidic experiments. At a given \mbox{$\text{Re}$}, healthy RBCs form a larger steady-state CFL than rigid cells, as indicated by the black dashed horizontal lines in Fig.~\ref{FIG_SIM}a. The same behavior is observed for the other investigated \mbox{$\text{Re}$}, as shown in Fig.~S6 in the Supplementary Material. The distance \mbox{$x_s$} and the time \mbox{$t_s=x_s/u_{\text{max}}$} that are required to reach the steady-state behavior in the numerical simulations are shown in Fig.~\ref{FIG_SIM}b as a function of \mbox{$\text{Re}$}. Here, we define \mbox{$x_s$} when 95\% of the value for the mean steady-state CFL is reached. We find that at \mbox{$\text{Re}=12$}, healthy RBCs reach the plateau CFL already around \mbox{$0.2\times L_\text{D}$}, while rigid cells require a longer length of approximately \mbox{$0.25\times L_\text{D}$} to achieve their steady-state CFL. Moreover, \mbox{$x_s$} does not systematically depend on \mbox{$\text{Re}$} for both healthy and rigid RBCs. The time \mbox{$t_s$} to reach the CFL plateau decreases with increasing \mbox{$\text{Re}$}, as shown in the lower panel of Fig.~\ref{FIG_SIM}b.

The CFL for the experiment and the steady-state CFL in the simulations are shown in Fig.~\ref{FIG_SIM}c for different \mbox{$\text{Re}$}. Here, we plot the upper CFL for the microfluidic experiments at the end of the daughter vessel (see Fig.\ref{FIG_CFLD}), which corresponds to the formation of a symmetric CFL. In Fig.~\ref{FIG_SIM}c, we find that the CFL in the simulations does not significantly depend on \mbox{$\text{Re}$} within the investigated range for both healthy and rigid RBCs. In the daughter vessel, we observe smaller CFLs for rigid than for healthy cells in the simulations, in agreement with our microfluidic experiments. Further, the behavior of CFL with \mbox{$\text{Re}$} is similar to the experimental observations in the mother channel (see Fig.~\ref{FIG_lowHt}b). Overall, the CFL determined in the numerical simulations is slightly higher by roughly \mbox{$\unit[1-2]{\um}$} compared to the experiments, similar to other numerical and microfluidic studies~\cite{Zhou2020a}. These differences could be due to the correction that is applied to determine the RBC surface based on the center of mass from the numerics. Another factor that could contribute to these differences could arise from a slight misalignment of the $x-y-$-plane of the microfluidic chip with the optical axes of the microscope. Due to the relatively deep channel (\mbox{$H=\unit[54]{\um}$}), having the chip slightly tilted can lead to an underestimation of the CFL. 

\section{Discussion}
Flowing through the inlet mother channel, RBCs are focused to specific positions along the channel width for healthy and rigid RBCs, as shown in Fig.~\ref{FIG_lowHt}a. For healthy RBCs, the emergence of two pronounced peaks close to the walls of the channel at the end of the mother channels is in contrast to the idea of an RBC core flow near the channel center, accompanied by a CFL near the channel walls.\cite{Cokelet1991, Secomb2017} Non-uniform RBC distributions in straight channels have previously been reported. Zhou~\etal~\cite{Zhou2020a} observed so-called off-center two-peak profiles for healthy RBCs in straight channels with dimensions on the same order of magnitude as the mother channel in this study. The authors argue that the emergence of these off-center two-peak profiles is co-determined by the spatial decay of hydrodynamic lift and the global deficiency of RBC dispersion in dilute suspensions.\cite{Zhou2020a} The positions of the peaks move towards the channel centerline along the flow direction until they stabilized around \mbox{$y/W\approx \pm 0.4 $}, similar to the peak positions observed in Fig.~\ref{FIG_lowHt}a at the end of the mother channel. However, in their study, these peculiar RBC distributions emerged at negligible inertia (\mbox{$\text{Re}<2\times10^{-4}$}).\cite{Zhou2020a} For dilute suspensions with hematocrits \mbox{$<\unit[1]{\%}$} shown in Fig.~\ref{FIG_lowHt} at \mbox{$\text{Re}>1$}, focusing effects can severely impact the spatiotemporal distribution of particles and cells in tubes and microchannels.

In general, lateral migration and focusing of particles in circular pipes due to inertial lift forces was first reported by Segr{\'{e}} and Silberberg.\cite{Segre1961} Inertial focusing is mainly driven by two dominating opposing forces. On one hand, the wall-induced lift force, due to the interaction between the particle and the channel wall, pushes the particle away from the wall. On the other hand, the shear gradient-induced lift force, due to the curvature of the velocity profile in the channel, directs the particle away from the channel center.\cite{DiCarlo2009, Amini2014a} These forces drive particles and cells towards specific equilibrium positions in the channel cross-section, which depend on the channel geometry, particle size and rigidity, and flow velocity. In square or rectangular microchannels, focusing of particles or cells is achieved for particle Reynolds number \mbox{$\text{Re}_\text{p}$} of the order of equal or larger than one, where \mbox{$\text{Re}_\text{p}=\text{Re}(a/D_{\mathrm{h}})^2$} with the particle or cell size \mbox{$a$}. Further, a certain length \mbox{$L_{\text{f}}$} is required for particles to reach stable equilibrium positions along the flow direction.\cite{Amini2014a} For rectangular channels, different stable focusing positions located near the faces of the channel exist. Depending on the aspect ratio of the channel cross-section, the distance from the inlet, and \mbox{$\text{Re}$}, particles and cells can be focused towards the wide channel faces or both the short and wide faces.\cite{Hur2010, Ciftlik2013, Zhou2013, Geislinger2014, Tanaka2022, Sugihara2021, Oh2022} For the experiments shown in Fig.~\ref{FIG_lowHt}, and with the RBC diameter \mbox{$a=\unit[8]{\um}$}, we find \mbox{$\text{Re}_\text{p}\approx0.24$}, \mbox{$L_{\text{f}}\approx\unit[17]{mm}$}, and \mbox{$\text{Re}=24$}. Although the \mbox{$\text{Re}_\text{p}$} is smaller than one, we expect inertial lift forces to partially affect RBC ordering and influence the observed distributions of healthy RBCs at \mbox{$\text{Re}>1$}. The lift force acting on the RBCs and hence, the focusing and separation of cells, further depend on their shape.\cite{Masaeli2012, Lanotte2016, Chen2017, Bazaz2020} In our study, the RBCs are rigidified in stasis where the cells exhibit a biconcave disk-like shape. While the healthy RBCs are able to adapt their shape according to the flow conditions in the channel, rigid RBCs retain their shape fixed in stasis. Hence, the difference in the distribution of rigid RBCs shown in Fig.~\ref{FIG_lowHt}a, emerges as a consequence of their impaired deformability, which results in different shapes in flow, in agreement with studies on particles and ellipsoids\cite{Masaeli2012} and rigid RBCs.\cite{Chen2017} Furthermore, Shen~\etal~\cite{Shen2016a} investigated the effect of RBC deformability on the inversion of hematocrit partition at microfluidic bifurcations. At low RBC concentrations, they observed an inversion of the hematocrit profiles that we observe for healthy and rigid cells. However, instead of using GA to irreversibly fix the RBC shape, the authors used different dextran concentrations to tune the viscosity contrast between the inner cytosol and surrounding fluid. This changes the dynamics of lift and hydrodynamic interactions of cells, as well as their transient shape dynamics,\cite{Recktenwald2021d} and might therefore lead to contradictory behavior with respect to the results reported in Fig.~\ref{FIG_lowHt}a.

At the bifurcation at the end of the mother channel, we observe RBC partitioning, which is affected by the asymmetry ratio of the T-channel, as shown in Fig.~\ref{FIG_lowHt}c and d, as well as distinct distributions for healthy and rigid RBC, shown in Fig.~\ref{FIG_lowHt}e. Early theoretical model predictions\cite{Chien1985} and experiments\cite{Fenton1985} on microvascular bifurcations revealed a nonlinear relation between the fractional RBC flux and the fractional bulk flow rate at \mbox{$\text{Re}\ll 1$}. The classical Zweifach-Fung effect states that the daughter vessel with the highest flow rate will collect a higher hematocrit.\cite{Fung1971} One simple rationalization is that the RBCs focus on the streamlines with a higher flow rate, while the CFL is formed and contains the streamlines with lower flow rates. The daughter vessel with the highest flow rate then collects blood enriched in RBCs, while the other collects a bigger proportion of blood depleted from RBCs. The focusing and characteristics of the CFL are important parameters influencing the partitioning of RBCs at bifurcations. Recent studies highlighted that the asymmetry of cell focusing observed after a bifurcation can significantly modify the partitioning of RBCs in subsequent bifurcations.\cite{Shen2016a, Mantegazza2020, Zhou2021, Merlo2022} Additionally, it has been shown that at low hematocrit, RBC flow tends to deviate from the Zweifach-Fung effect and that deformability plays an important role in this reverse behavior of partitioning.\cite{Shen2016a}

Additionally to our results on the RBC focusing phenomenon, we provide further insight into the CFL formation along the mother and daughter channels and their dependency on the applied pressure drop, hematocrit, and cell rigidity, as highlighted in Fig.~\ref{FIG_lowHt} and Fig.~\ref{FIG_CFLD}. The formation of a CFL is attributed to the tendency of RBCs to laterally migrate away from the vessel wall. The main driving forces for this migration include size exclusion effects due to the finite size of the RBC, boundary interactions, and macromolecular layer exclusion effects due to the inner lining of the vessel, and the curvature of the velocity profile in the vessel that produces a tendency for migration toward the vessel centerline.\cite{Kim2009, Pries2008, Katanov2015, Secomb2017} As shown in Fig.~\ref{FIG_lowHt}a and Fig.~\ref{FIG_CFLD},  we do not observe a CFL at the beginning of the mother channel and similarly, at the beginning of the daughter vessel for the down CFL.

For straight microfluidic channels, the CFL growth along the flow direction was investigated before. Zhou~\etal~\cite{Zhou2020a} recently investigated the CFL growth in a straight rectangular channel comparable to the dimensions of the mother channel used in this study, however, at negligible inertia \mbox{$\text{Re} \ll 1$} using numerical simulations and microfluidic experiments. At low RBC concentrations, they observed a build-up of the CFL that followed a power-law behavior with exponents between 0.26 and 0.4. The CFL in their simulations increased over a length of \mbox{$28\times L/D_{\mathrm{h}}$} without saturation. In their microfluidic experiments, the CFL growth occurred over more than \mbox{$46D_{\mathrm{h}}$} without reaching an equilibrium for RBC concentrations of \mbox{$\unit[1]{\%}$}. In our study at \mbox{$\text{Re}>1$}, the RBC travel over roughly \mbox{$350\times L/D_{\mathrm{h}}$} at the end of the daughter vessel showing both saturation and recovery of symmetry. Based on the evolution of the lower CFL in the daughter vessel in the experiments, shown in Fig.~\ref{FIG_CFLD}, we consider the CFL to have reached a steady state at \mbox{$x_{\text{D}}=L_{\text{D}}/2$} (\mbox{$\approx175 D_{\mathrm{h}}$}) for \mbox{$\unit[0.1]{\%Ht}$}. Our numerical simulations further show that the length to reach the steady-state CFL is not significantly influenced by \mbox{$\text{Re}$}. However, the simulations also demonstrate that healthy RBCs achieve their equilibrium CFL under flow faster than rigid cells. For higher RBC concentrations in the experiments, the CFL saturates even faster, as shown in Fig.~S5 in the Supplementary Material. This is due to enhanced RBC interactions and collisions in the RBC core flow that increase with the hematocrit and that push the cells toward the channel walls.

Besides this dependency of the CFL growth on the RBC concentration, the magnitude of the CFL decreases with increasing hematocrit. Additionally, we find that the CFL thickness is always smaller for rigid RBCs compared to healthy RBCs under the same experimental conditions (\mbox{$\text{Re}$} and hematocrit) and at the same position (\mbox{$x_{\text{D}}$} or \mbox{$x_{\text{M}}$}), in agreement with previous studies.\cite{Fujiwara2009, Bento2018, Bento2019, Abay2020} In bifurcating microfluidic channels, the partitioning and flow of RBC suspensions and the emergent CFL revealed heterogeneous RBC distributions, skewed and blunt velocity profiles, and an enhancement of the thickness of the CFL at higher hematocrit.\cite{Sherwood2012a, Sherwood2014, Sherwood2014a, Kaliviotis2017, Kaliviotis2017a, Kaliviotis2018} Yamamoto~\etal~\cite{Yamamoto2020} studied the partitioning of RBCs through asymmetric bifurcating microchannels at \mbox{$\unit[0.5]{\%Ht}$}. They observed that with an increasing fractional flow rate in the daughter vessels, the CFL ratio between the mother and the daughter vessel decreases, similar to our results shown in Fig.~\ref{FIG_lowHt}d. While other numerical studies already showed that rigidification of RBCs tends to decrease the CFL width,\cite{Yin2013, Zhang2009} those previous studies do not provide a comparison with experimental data as our study does. This is particularly relevant since most of the combined studies showed some discrepancies between simulations and experiments.\cite{Zhou2020a, Fedosov2010}

Understanding the distribution profile and CFL development in bifurcation channels is paramount to understanding the flow and RBC in vascular networks. In such network structures, which often consist of multiple series-connected bifurcations, the CFL in successive bifurcations is crucially impacted by the RBC distribution in the previous bifurcation.\cite{Bento2018, Balogh2019, Bento2019, Li2022a} Zhou~\etal~\cite{Zhou2021} found that changing the fractional flow rate between the daughter vessels leads to a complete depletion of RBCs in one branch of the successive bifurcation. Similarly, the distinct RBC distributions of healthy and rigid RBCs in the different ROIs in the daughter vessels (see Fig.~\ref{FIG_lowHt}e) would lead to dramatically different RBC organizations in subsequent branching channels. Namely, the formation of a pronounced peak of rigid cell concentration close to the lower daughter walls is a potential mechanism through which rigid cells can perturb the healthy flow of RBCs in the circulatory network.

\section{Conclusions}
Studying the effect of RBC rigidity in microfluidic flows advances our knowledge of blood flow in vivo and is crucial to understand the impact of pathological RBC changes on their flow properties, especially in complex geometries, such as bifurcating vessels and networks. In this study, we performed microfluidic measurements at various positions along the channel flow direction in a bifurcating microchannel to understand the RBC flow behavior, their lateral organization across the channel width, and the CFL phenomenon covering a broad hematocrit and Re range of \mbox{$\unit [0.1-5]{\%Ht}$} and \mbox{$\text{Re}=6-24$}, respectively.

In the mother channel, we observe different RBC focusing patterns that result in the emergence of two peaks close to the channel walls for healthy cells, while rigid RBCs predominantly flow along the channel centerline. Arriving at the bifurcation, these differences lead to distinct RBC distributions in the daughter vessels for rigid and healthy RBCs, which persist until the end of the channel. Our microfluidic in vitro experiments on artificially hardened RBC with GA show that the partition at the level of the bifurcation depends strongly on RBC deformability. Further, we reveal how the bifurcation affects the development of the different CFLs in the daughter branches. Our numerical simulations further demonstrate that the length to reach the steady-state CFL depends on cell rigidity. Since the distribution of RBCs after a bifurcation has been shown to be a determining factor for the distribution of RBCs in the microcirculatory network,\cite{Zhou2021} the higher concentration close to the lower daughter walls demonstrated by our study for rigid cells is a potential mechanism through which blood flow can be modified by pathologically stiffened cells. Our findings highlight the importance of understanding the influence of RBC rigidity on partitioning in complex flow fields, and thus oxygen delivery in the microcirculatory network.

\begin{acknowledgments}
This research was funded by the Deutsche Forschungsgemeinschaft (DFG, German
Research Foundation) – project number 349558021 (WA~1336/13-1, HA~4382/8-1, and RE~5025/1-2). Y.R. acknowledges funding by the Marie Skłodowska-Curie grant agreement No. 860436—EVIDENCE. A.D. acknowledges funding from the Young Investigator Grant of Saarland University.

\end{acknowledgments}

\bibliography{main}

\end{document}